\documentclass{article}

\usepackage{arxiv}

\usepackage[utf8]{inputenc} 
\usepackage[T1]{fontenc}    
\usepackage{hyperref}       
\usepackage{url}            
\usepackage{booktabs}       
\usepackage{amsfonts}       
\usepackage{nicefrac}       
\usepackage{microtype}      
\usepackage{lipsum}		
\usepackage{graphicx}
\usepackage{natbib}
\usepackage{doi}
\usepackage{hyperref}       
\usepackage{url}            
\usepackage{booktabs}       
\usepackage{amsfonts}       
\usepackage{nicefrac}       
\usepackage{microtype}      
\usepackage{lipsum}		
\usepackage{graphicx}
\usepackage{natbib}
\usepackage{doi}
\usepackage{graphicx}

\title{BIOPAK FLASHER: Epidemic Disease Monitoring and detection in Pakistan using Text Mining}

\author{\author{Muhammad Nasir$^{1}$  \and
        Maheen Bakhtyar$^{1}$ \and
        Junaid Baber$^{1}$ \and
        Sadia Lakho$^{1}$ \and
        Bilal Ahmed$^{2}$ \and 
        Waheed Noor$^{3}$
}
\thanks{Use footnote for providing further
		information about author (webpage, alternative
		address)---\emph{not} for acknowledging funding agencies.} \\
\And
{\includegraphics[scale=0.06]{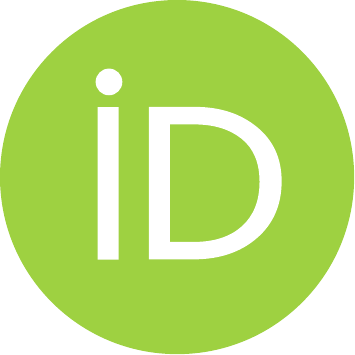}\hspace{1mm}Muhammad Nasir} \\
	Department of CS and IT\\
	University of Balochistan\\	
	\texttt{muhammadnasirsharif11@gmail.com} \\
	\And
	{\includegraphics[scale=0.06]{orcid.pdf}\hspace{1mm} Maheen Bakhtyar} \\
	Department of CS and IT\\
	University of Balochistan\\
	\texttt{maheen.bakhtyar@um.uob.edu.pk} \\
\And
{\includegraphics[scale=0.06]{orcid.pdf}\hspace{1mm}Junaid Baber} \\
	Department of CS and IT\\
	University of Balochistan\\	
	\texttt{junaidbaber@ieee.org} \\
	\And
	{\includegraphics[scale=0.06]{orcid.pdf}\hspace{1mm} Sadia Lakho} \\
	Department of CS and IT\\
	SBK Women's University Balochistan\\
	\texttt{sadia.lakho@gmail.com} \\
\And
	{\includegraphics[scale=0.06]{orcid.pdf}\hspace{1mm} Bilal Ahmed} \\
	Department of CS and IT\\
	University of Balochistan\\	
	\texttt{bilal788@gmail.com} \\
\And
{\includegraphics[scale=0.06]{orcid.pdf}\hspace{1mm}Waheed Noor} \\
	Institute of Management Sciences\\
	University of Balochistan\\	
	\texttt{a.waheed.noor@gmail.com} \\
}

\renewcommand{\shorttitle}{\textit{arXiv} Template}

\hypersetup{
pdftitle={A template for the arxiv style},
pdfsubject={q-bio.NC, q-bio.QM},
pdfauthor={David S.~Hippocampus, Elias D.~Striatum},
pdfkeywords={First keyword, Second keyword, More},
}

\begin{document}
\maketitle

\begin{abstract}
	Infectious disease outbreak has a significant impact on morbidity, mortality and can cause economic instability of many countries.  As global trade is growing, goods and individuals are expected to travel across the border, an infected epidemic area carrier can pose a great danger to his hostile.  If a disease outbreak is recognized promptly, then commercial products and travelers (traders/visitors) will be effectively vaccinated, and therefore the disease stopped.  Early detection of outbreaks plays an important role here, and beware of the rapid implementation of control measures by citizens, public health organizations, and government. Many indicators have valuable information, such as online news sources (RSS) and social media sources (Twitter, Facebook) that can be used, but are unstructured and bulky, to extract information about disease outbreaks.  Few early warning outbreak systems exist with some limitation of linguistic (Urdu) and covering areas (Pakistan).  In Pakistan, few channels are published the outbreak news in Urdu or English. The aim is to procure information from Pakistan's English and Urdu news channels and then investigate process, integrate, and visualize the disease epidemic.  Urdu ontology is not existed before to match extracted diseases, so we also build that ontology of disease.
\end{abstract}


\section{Introduction}
In past many infectious diseases hampered humankind, like the Whooping cough outbreak in U.S. during the summer of 2012, Periodic outbreaks of H7N9 in China during 2013-2014 and emerging MERS outbreak in Saudi Arabia, Severe Acute Respiratory Syndrome (SARS), West Nile Virus and drug resistant tuberculosis \cite{zhao2012national}. Diseases are spread exponentially within a population. Peoples and products travel across the border in high volume so that an infection can rapidly attain a global scale.\\
A pandemic HiNi (“Swine Flu”) appeared in 2009 \cite{szomszor2011twitter}. To increase this outbreak, a globally open online web-based framework gave warning to everyone concerned.  The disease was catered for in response to this warning, and public understanding of emerging infection has increased. Online web-based early warning system’s traffic rate was increased about 100 \% in 2-3 days. It gives birth to online early warning globally infection awareness. Hence, globally feels that there must be a system that alerts the early warning infectious disease timely vaccinate and cure the victims. Presently, nearly all major outbreaks, investigated by the World Health Organization (WHO) are ﬁrst identiﬁed through these informal online sources.

 The emergence of ubiquitous internet service and build up different internet-based resources like blogs, online newspapers (Google News), social media (Twitter), and discussion forums shape the world dynamic fast and very efficiently. Such services are highly useful in helping to diagnose, assess, incorporate and forecast outbreaks of infectious diseases or diseases spreading to new regions. Intelligent modeling of such sources using text-mining methods such as topic models, deep learning and dependency parsing can lead to automated generation of the mentioned surveillance tools..\\
The above resources produced big data at very high speed, and the quantity of that data is increasing day by day. Despite big data problem, these resources have the unstructured type of data, which is difficult to process, and extract useful information. Reading and assimilating a large number of reports daily basis is a hectic and burdensome task.  .\\
 . With BIO PAK FLASHER, a web-based solution focusing primarily on URDU \cite{jabbar2016analysis}, Pakistani Urdu news outlets, and covering Pakistani rulers and urban areas, we plan to solve these challenges with the automated framework for extraction, analysis, query, screening, incorporation, and visualization of unstructured and bulky disease outbreaks.  Our system collects outbreak data from many sources such as news through the Google News aggregator, Pakistani local news channels published in Urdu, expert-curetted accounts such as ProMED \cite{morse1996promed} Mail, and validated official alerts such as World Health Organization announcements. By using the text processing algorithms, the system classifies outbreak in location and disease, after that overlay such data on an interactive geographic map. We measure the accuracy of the classification algorithms based on the level of human curation necessary to correct misclassiﬁcation and examine geographic coverage.  With state-of-the-art NLP technology, there is a need to compare all Urdu terms with a disease database called ontology to figure out the disease when the sentence is tokenized and all POS deleted. There is no disease ontology \cite{schriml2012disease} exist before to do such work. Here we also build Urdu-based Ontology with different patterns for further reference. .\\
 Google Maps API consists of a set of classes within a JavaScript container that opens within an XHTML page. Each time the Google Maps web page opened, these classes are loaded from Google, and it has embedded JavaScript object on which all the functionality of the Google Maps is based. To visualize outbreaks, we use Google Maps.BioPak Flasher has a user- friendly interface and providing important disease outbreak information with text processing algorithms and also referred user to view full article with web link integration.
The rest of the paper is composed of Section 2 Literature review presents the background of infectious disease outbreaks and basic terminologies used in this research. Section 3 is about the Methodology used in this research, Section 4 Describes the implementation and Analysis of the proposed methodology. Section 5 presents the conclusion and future work for this research.

\section{Literature Review}

There are few traditional disease surveillance systems like Centers for Disease Control (CDC) \cite{berrios2017centers}, ProMED \cite{morse1996promed}, which released data of infectious disease outbreaks with delay due to their implemented setup (take news mail, get expert opinion then release outbreak announcement).  In the modern period, the epidemic must be published in real-time. To cater to evolving diseases, take control steps quickly, and avoid them from spreading further, an early warning/detection of the outbreak system is important. These surveillance systems have some limitations. These programs do not protect all languages, so the epidemic news reported in state and local news would be missed. The system must be capable to process these languages so that any news published in theses languages will also be covered to give authenticated real-time system. Urdu is the national language of Pakistan and speaks in many other countries as well in non-official manners. Urdu is also not covered by these systems. Many local publishers like VOA-Urdu, Geo, Express, Jang, etc. published their news in Urdu. Moreover, these systems have some geographical limitations. These systems are not covering the ruler/ urban areas of some countries, especially Pakistan.\\
Infectious disease is increasing day by day due to unhygienic environment, people are in close contact with livestock or wildlife and chemicals, etc. Infectious diseases lead to mortality so awareness is a very essential part of the current era. Delayed detection of the outbreak may cause some mortality and may be created hinders in controlling and eradication process. Such mortality rate can be minimized if timely recognized. If we know in real- time, then the control and eradication measures also taken time to reduce mortality rate. \\
Moreover, when we want to visit somewhere or required to do trade with some area, we must consider emerging pathogens that causes the epidemic. Shifting of emerging product or human can cause the spreading of that epidemic in the hosting area the traditional outbreak system has a manual alarm that often requires a delay, but our, alert system, swift, and covering Pakistani news is our disease outbreak system written in Urdu. This system also helps in general awareness of infectious outbreaks and its remedies measures timely.

\subsection{Epidemic Breakdowns In Pakistan}

A government of Pakistan Official Body releases the statistics in 44th dated Mar 2019 to June 2019. According to these, Figure 1 alerts has been generated in the country, Figure 2 shows the statistical results of diseases outbreaks in Pakistan. 

\subsubsection{Epidemic (A Communicate-able Disease) }

An epidemic \cite{newman2002spread} occurs when an infectious disease spreads rapidly in contact with too many people. An epidemic is an outbreak of a disease among members of a specific population that exceeds the extent of occurrence of the disease normally found in that population. Epidemics affect those members of the population who do not have an acquired or inherent immunity to the disease. Although infectious organisms cause most epidemics, the term can be applied to an outbreak of any chronic disease, such as lung cancer or heart disease. For example, in 2003, the severe acute respiratory syndrome (SARS) epidemic took the lives of nearly 800 people worldwide. Statistics and big outbreaks in history with casualty.

\subsubsection{RSS Feed:  }
RSS stands for "Really Simple Syndication". It is a way to easily distribute a list of headlines, update notices, and sometimes content to a wide number of people \cite{johnson2009developing}. Computer programs that organize those headlines and notices for easy reading use it. Most people are interested in many websites whose content changes on an unpredictable schedule. News sites, community, and religious group information pages, product information pages, medical websites, and weblogs are examples of such websites. Repeatedly checking each website to see if there is any new content can be very tedious.
\begin{figure}
	\centering
 \includegraphics[width=0.75\textwidth]{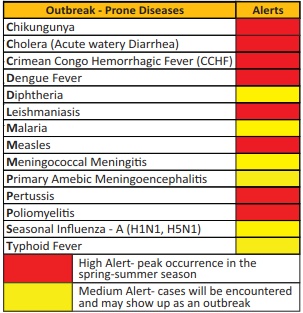}	
	\caption{Recent tutbreaks in Pakistan.}
	\label{fig:fig1}
\end{figure}

\begin{figure}
	\centering
 \includegraphics[width=0.75\textwidth]{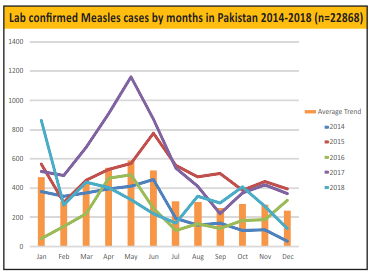}	
	\caption{Statistics of different disease outbreak in Pakistan.}
	\label{fig:fig2}
\end{figure}

\subsubsection{How It Works:}
RSS works by having the website author maintain a list of notifications on their web-site in a standard way. This list of notifications is called an "RSS Feed". People who are interested in finding out the latest headlines or changes can check this list. Special computer programs called "RSS aggregators" have been developed that automatical-ly access the RSS feeds of websites you care about on your behalf and organize the results for you. (RSS feeds and aggregators are sometimes called "RSS Channels" and "RSS Readers".)
\subsubsection{RSS Feed Structure: }
An XML File provides very basic information to do its notification. It is made up of a list of items presented in order from newest to oldest. Each item usually consists of a simple title describing the item along with a more complete description and a link to a web page with the actual information being described. Sometimes this description is the full information you want to read (such as the content of a weblog post) and sometimes it is just a summary.

\section{Methodology}
The different aspects of this research including data collection were the compulsory requirement for further utilization of the data. The essentials tasks are explained figure 3.\\
In this figure, we presented the layer view of our model. We preferred to work in layer because it is easy to wok easily if we have to do any changing. We divide our model into four layers; each layer is dependent on its precedence layer. Each layer work is defined clearly.  The first layer is the Data acquisition layer, which retrieves data from the RSS, feed and puts it into the database for further processing.  The second layer is the Data filtering and preprocessing layer in which we eliminate and filter out unnecessary diacritics and punctuations then performing some NLP (Natu-ral Language Processing) techniques like tokenization, stemming, and stop words removal of Urdu language. Then our data has epidemic, city, date, and some garbage or unnecessary data. The third layer of Event Recognition is designed to find out the epidemic disease and concerned area with latitude and longitude. Our last layer is the Visualization layer, which deals with visualization of epidemic outbreak in the con-cerned area will show on Google map with marker view and store events in database for displaying it next 90 days. User can also view source evidence by simply clicking on a pinpoint map as well as associated headlines for the event. 
\begin{figure}
	\centering
 \includegraphics[width=0.75\textwidth]{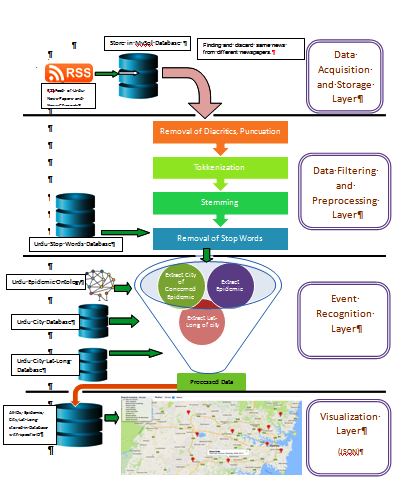}	
	\caption{Abstract overview of the BioPak flasher.}
	\label{fig:fig3}
\end{figure}
\subsection{Data Acquisition and Storage Layer: }
In this layer. We acquire the data from Pakistani news channels and newspapers that are offering online resources (RSS feed) in Urdu and store these data into a database for further processing. Many online actors can provide news like RSS, tweeter, LinkedIn, Facebook, etc. However, our choice is RSS feed. RSS stand for Really Simple Syndication or sometimes refer as Rich Site Summary. Actually, this technol-ogy was introduced to avoid bulky and uninterested news, which is being pushed by a provider. This is also used to keep updated user with upcoming news, blogs, financial information, health, sports, classified sites, and government alerts. RSS feed is written in XML code which is easy to process by computer. It is a structured data type. 
 Figure 4 shows a sample of newsreader who acquired the news from Voice of America in Urdu. We can select any topic in which we are interested, in our case health topic is most prominent and maintained by almost every RSS feed provider. So we subscribe to the health RSS feed.
	When we click on any news feed, it will refer us to its description web page shows in Figure 5, where we can see all detail of the news. In this figure, we refer to the voice of America page where we can find the detail of said news.
\begin{figure}
	\centering
 \includegraphics[width=0.75\textwidth]{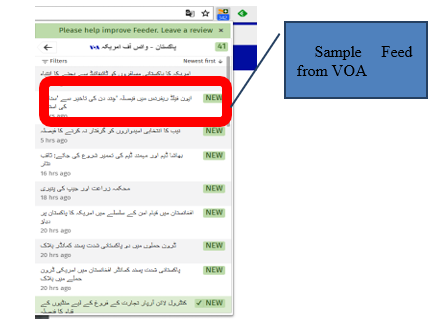}	
	\caption{Sample overview of RSS feeds.}
	\label{fig:fig5}
\end{figure}

\begin{figure}
	\centering
 \includegraphics[width=0.75\textwidth]{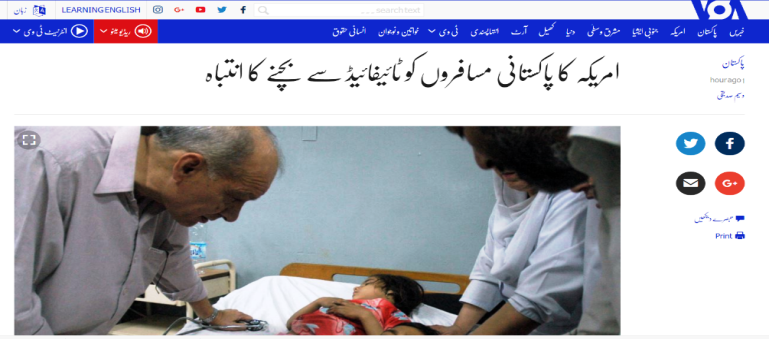}	
	\caption{Sample screenshot of the News.}
	\label{fig:fig5}
\end{figure}
Our focus of research covered Pakistan and its national language Urdu Pakistan has several outlets that include printed media, news channels, several online podcasts, and internet-based TV channels used for consumer information. These all channels are published in different languages according to a location like English, Urdu, Sindhi, etc. Therefore, Urdu RSS feed newspaper or channels are our main actor for getting information. However, very few channels are maintained their RSS feed in Urdu. Because our research is focused on the RSS mechanisms so that we only focus that news, channels, which have their RSS, feed in Urdu and store it.

\subsubsection{Data filtering and Pre-Processing: }

There are many types of data which is published in the RSS feed. Input news items are not considered to be cleaned so making them ready for further processing is im-portant. Raw data, if obtained from the web normally includes HTML and or XML tags so they should be removed for further processing. In addition, text format (utf-8) for special characters is important so that no error occurs while processing.

\subsubsection{Removal of Diacritics: }
The Urdu language is using diacritics to clarify the pronunciation of Urdu words. Commonly used diacritics are  Zabar, Zer, Pesh, Jazam, and shad, etc. Our raw data received from the database has many unnecessary elements that are required to be filtered shows in figure 6. In this stage of filtering, it is necessary that data must be filtered from diacritics. If it is not done then in processing, different result will be seen and our final output is very different from the actual. \\
\centering
 \includegraphics[width=0.55\textwidth]{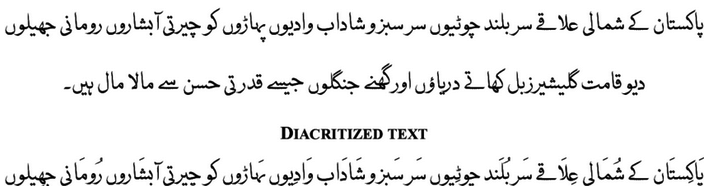}

\subsubsection{Removal of Punctuation}

Urdu language sentences have different punctuations, which used in the construction of sentences. These punctuations are undesired in our processing so that we removed these. Some punctuations which mostly used in the Urdu language are “ “ ” ‘’ – [](),.;: “

\subsubsection{Data Preparation (Tokenization with NLTK   Library)}

Urdu tokenization is not easy as English due to its structure. However, Unicode recommends using Zero Width Non-Joiner character in this context \cite[4]. There would be an alignment of tags with the terms present at the simple stage, so tokenization is also an essential aspect of sorting. The text will be tokenized into sentences and phrases into words in the process.

Tokenization is a very common feature of NLP. Tokenization is a process to break a text string into single entities that may include words, symbols, phrases, etc. In our case, we apply this technique to divide the string into words, e.g. 
\begin{figure}
	\centering
 \includegraphics[width=0.55\textwidth]{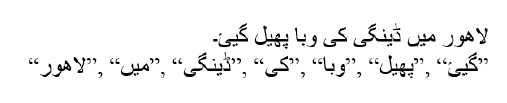}	
	
	\label{fig:fig7}
\end{figure}
Further processing is taken place on the resulted array after tokenization. 

\subsubsection{Building up of Urdu Stop Words (USW) Dictionary }
After tokenization, now many unwanted words have no meaning when we tokenize but these words have more worth when used in combination and clear the concept of sentences. Unfortunately, there is not much work done in Urdu, so we used stop word dictionary Stop Words Removal (Based on USW Dictionary)
These dictionaries are used as a reference for the removal of stop words. This is another technique of NLP, which are tags, or separates various parts of speech in the text. For filtering, stop words are also attached with this technique to filter indifferent data. Epidemic disease, city, and date of occurrences are needed to be extracted from the text string. Everything else in the string is irrelevant and should not be extracted or in other words, should be deleted from the string. 
Tokenization has been done in the previous step. Keeping in view the resulted string, now the irrelevant data should be filtered out from the string. At this stage, we have the tokenized string as:\\
\centering
 \includegraphics[width=0.55\textwidth]{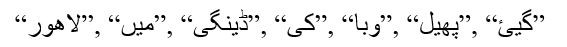}\\
After deletion of Stop word from string, we will be left with the following array:\\
\centering
 \includegraphics[width=0.55\textwidth]{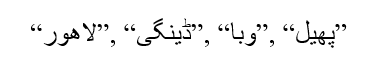}\\
Now the leftover data included only disease, city, and date or in some cases some garbage or irrelevant data.  
\subsubsection{Data Correction (Data Stemming on Tokenize word to their basic form)}
It is also known as lemmatization in which the word is changed to its very basic form by removing suffixes and will be converted to its root form. This makes the handling of words very easily. The significance of stem is very well explained e.g. \includegraphics[width=0.09\textwidth]{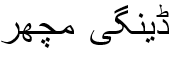}may be written as  \includegraphics[width=0.09\textwidth]{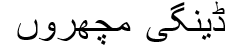} Failed to do so sometimes create problems in processing which ultimately affects our result. 

\subsubsection{Data Correction (Different Format of Single Word)}
Urdu is a versatile language and words can be written in different styles. Like \includegraphics[width=0.06\textwidth]{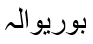} to \includegraphics[width=0.07\textwidth]{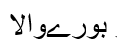},  \includegraphics[width=0.04\textwidth]{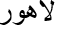} to \includegraphics[width=0.04\textwidth]{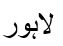}
in this case our dictionary is unable to find these types of word and failed to produce results. Then we refer these cases to our format pattern dictionary where we describe the different patterns of the same word. If a word is not recognized by “Urdu based epidemic and city” dictionaries then we will move to this dictionary for finding the diversion of said word. This dictionary is flexible and regularly updated on the feed-back and custom detection by users. If we checked every word in this dictionary and found that maximum words exist in  their standard shape, then it is useless and  resource-consuming that we checked every word for its reformatting. This dictionary is optional. 

\subsection{Event Recognition Layer}
After necessary filtering and preprocessing, we have now reduced the array of string. In this layer, we recognize the epidemic and also its occurrence city with the date stamp. This layer is also eliminating duplicate news from different channels and newspapers.

\subsubsection{Epidemic Directory / Ontology in Urdu (Creation and updating of said Directory / Ontology on regular basis)}
To find out disease from the preprocessed array, we must have a dictionary or data-base from where we can match and find the result. These types of dictionaries are available in English but in the Urdu language, it does not exist. We prepared URDU Directory / Ontology Dictionary which contains initially approx. 50 diseas-es/Pathogens/ Infections with its English version.  

\subsubsection{Event Recognition (Epidemic) }
In this stage, we extracted the disease from the news feed in connection with epidem-ic disease ontology. There may be some confusion regarding disease nomenclature like \includegraphics[width=0.03\textwidth]{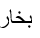} or \includegraphics[width=0.03\textwidth]{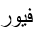} or fever, we removed this confusion with us reformat dictionary. We made them in simple and basic level, so that the exact disease is to be identified.

\subsubsection{City Database in Urdu (Enlisted the all cities of Pakistan in URDU) }
Pakistan has 374 cities. There is no database exist which describe the Pakistan cities names in Urdu, so for our purpose of city recognition, we will make a database. This database/ dictionary will use to detect the city of an epidemic outbreak. Urdu pub-lishers are not always following standard patterns and published their news in very random ways, sometimes the name of the city is written as \includegraphics[width=0.06\textwidth]{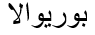} and sometimes \includegraphics[width=0.05\textwidth]{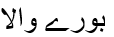} or sometimes “Burewala”. This versatility in city names creates confusion for the program to find out the exact city. We will make our dictionary/ database in such a way that caters all these problems. We put different patterns of a city in Urdu and English also. So, when querying this dictionary, it will return us the basic and standard format of the city. We will also make this database relaxant so that names of cities or their different formats may include later on. After determining the disease, it is neces-sary to find out the epidemic area so that we can visualize it on a map. 
\subsubsection{Entity Recognition (City) (Find City of concerned Epidemic)}
In the light of the city named dictionary, we extract city names from the RSS title. If the city is not identified in the RSS feed title, then we moved on to the description to find out the city. There may be variants of city names published in RSS feed that caters as described above. When we query the city database it will return our city name in that format which is recognized by our latitude and longitude stage. 

\subsubsection{City Lat-Long Database in Urdu (Enlisted Latitude and Longitude of Pakistan cities in Urdu) }
The world is divided into latitude and longitude. Latitude and Longitude are the units that represent the coordinates at the geographic coordinate system. To pinpoint any location or refer to this on the map, we must know the latitude and longitude of said area.  The map only understands the latitude and longitude. So that to visualize the city on the map we must know the latitude and longitude of said city. Before this, there is no database that describes the latitude and longitude of Pakistan cities in Urdu. We prepared a database that will return the latitude and longitude of each city of Pakistan when inquired. These are the fixed parameters and not changed with the passage of time so that we make it robust. 

\subsubsection{Recognition of City Lat-Long (Find Latitude and Longitude of Concerned City)}
After identification of city, we will extract its latitude and longitude from the Lat-Long database and will store this information in a database for further processing of JSON to visualize the epidemic on the map.

\subsection{Visualization Layer (JSON)}
In this layer, we visualize the epidemic outbreak \cite{newman2002spread} with the concerned area and also used the cluster marker to simplify the map vision which removed the so many pins clutter and combined the same area with a marker. Epidemic disease, city, city lat-long, date, and link are stored in a dedicated separate database. 

\subsubsection{Database Storage (All IDs, Epidemic, City, Lat-Long stored in Database with spe-cific ID)}
After extracting all necessary information regarding the disease, city, lat-long, date, link and system generated ID will be stored in a dedicated database.  This infor-mation is accessed by JSON and will be used to view on map, we will set this data-base frequency up to 90 days which will be reduced on user-customized query. 

\subsubsection{Web Site Backend (Fetch data from DB with respect to Location/ Disease)}
Website required certain types of data to be used on a map. These data included the epidemic disease and area with lat long. The Database is store in the list and  is  used in the visualization.

\subsubsection{Web site Backend Fetching Types:}
\begin{itemize}
	\item Fetch data from DB with respect to Location/ Disease
	\item Fetch customized data as per query: database can be customized as per user requirements.	
\end{itemize}

\subsubsection{Visualization (In term of pathogen, syndrome, or text type)}
For visualization purpose, we have used the JSON. First, we module our data into JSON pattern and then this data are used by map to visualize. Wherever the disease outbreak occurred, a marker is placed there and a short note will appear there. 

\subsubsection{Marker Clustering Google Map Marker Clustering (For more clarity in visuali-zation)}
If any place has many markers \cite{netek2019performance} then these markers are combined and displayed as a single marker, which simplified the view.

\subsubsection{Source Evidence (URL will open detail for  that epidemic)}
Source evidence is a feature to facilitate the user with full detail of the news. If some-one wants to know further details of the concerned epidemic, then it will just click and web refers him to the concerned page for more detail. Different colors are used to make it more comprehensive for visualization.

\section{Experiments \& Results}
Our developed model has four layers and each layer has different stages. Here we describe the output of each stage. 

\subsection{RSS Feed Output:}
We used almost all available resources that are being offered the Urdu RSS Feed. Here we only see the results of SUCHTV news. We automate it to run after the calendar day has been changed. SUCHTV news almost issued 10-20 RSS feed daily when the calendar date changed. We also tune our system so that it will not miss any feed. If we want to run manually or mistakenly run the developed system then it captured the RSS feed but did not put it in the database because there is already that feed available with access No control. Figure 7 shows the RSS feed which is being captured by our system. 
\begin{figure}
	\centering
 \includegraphics[width=0.75\textwidth]{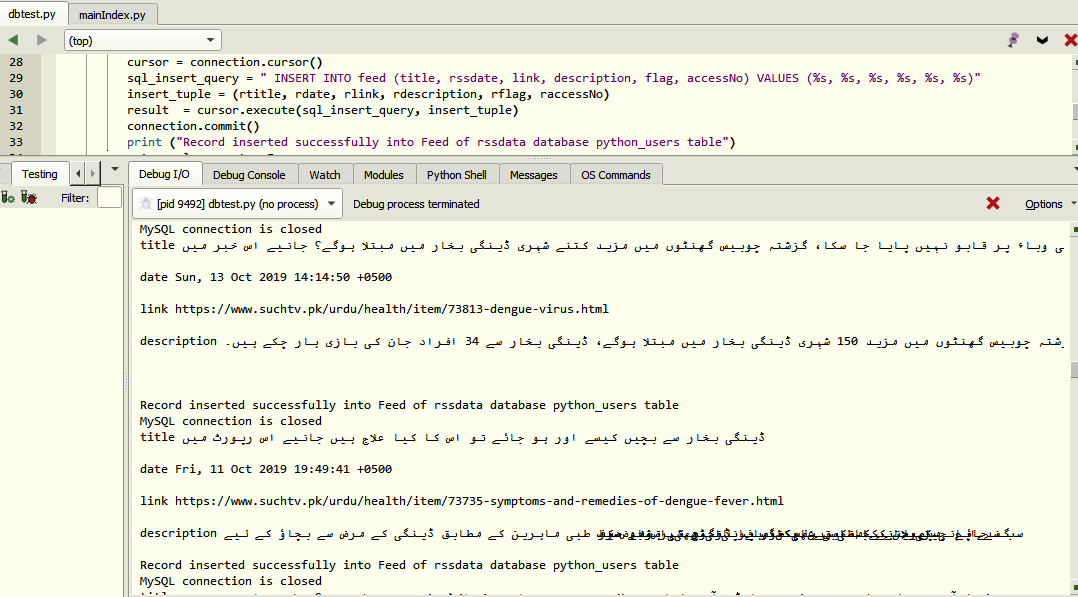}	
	\caption{SACH TV News RSS Feed.}
	\label{fig:fig6}
\end{figure}
\subsection{RSS Feed in Database}

Here we captured the title, date of news which is being provided by RSS provider (SUCHTV news), description, access No, and flag. The flag is placed here to lessen the burden of the system. When our system placed the new RSS feed in the database then the flag is set to “1”. But when the system processed that news and found the results then it assigned it “0” so that next time when the system approaches the data-base then only a new RSS feed will be processed, hence we less the burden of the system. Figure 8 shows the RSS feeds are placed in the database. 

\begin{figure}
	\centering
 \includegraphics[width=0.75\textwidth]{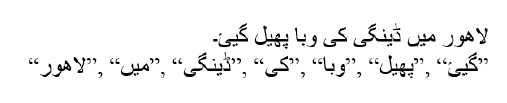}	
	\caption{RSS Feed stored in Database.}
	\label{fig:fig7}
\end{figure}
\subsection{Epidemic Outbreak Detection }

The accuracy of our system is depending on the detection of epidemic diseases from every RSS feed. For this purpose, we maintained and directory of epidemic disease as reference because Urdu based ontology is not exist before. Figures 9 shows the detected disease from RSS feed at system run time, figure 10 show the summary of detected disease with their database ID for easy tracking. 

\begin{figure}
	\centering
 \includegraphics[width=0.75\textwidth]{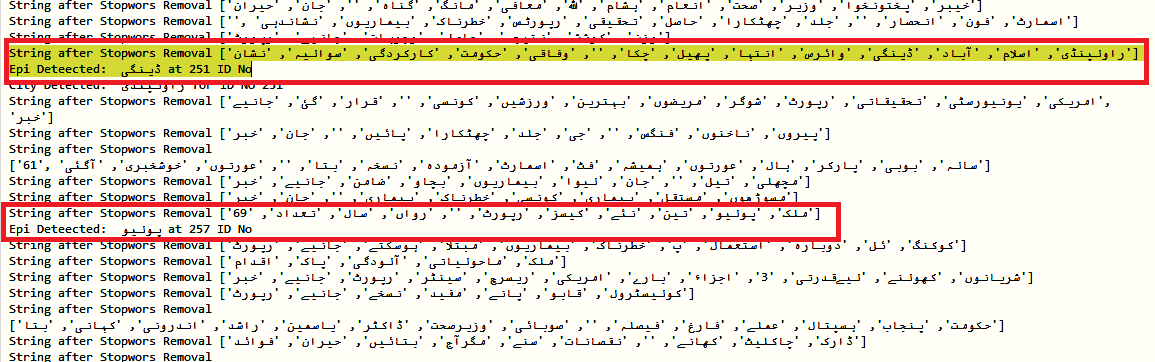}	
	\caption{Run time Epidemic Detection.}
	\label{fig:fig8}
\end{figure}

\begin{figure}
	\centering
 \includegraphics[width=0.75\textwidth]{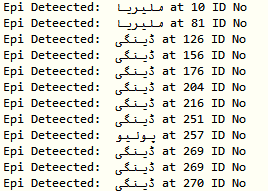}	
	\caption{Summary of detected disease with database ID.}
	\label{fig:fig9}
\end{figure}

\subsection{City Detection and city latitude City Latitude and Longitude Detection:}
Figure 11 shows the latitude and longitude of the detected city.
\begin{figure}
	\centering
 \includegraphics[width=0.75\textwidth]{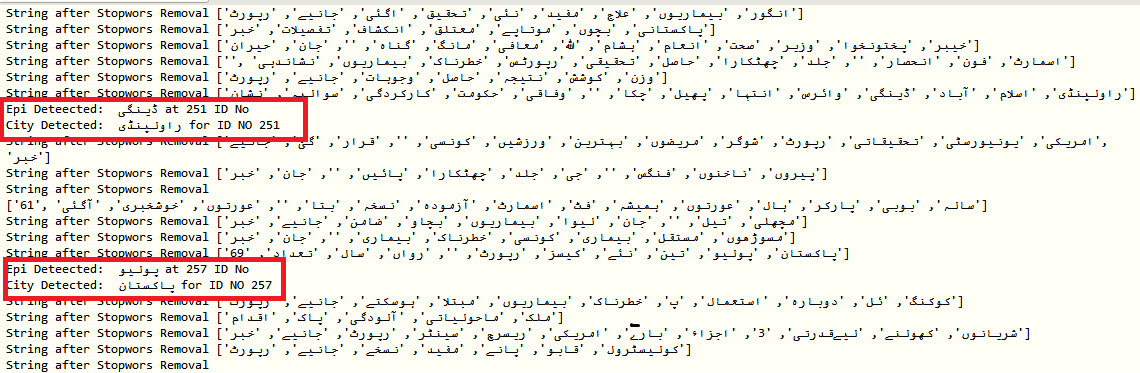}	
	\caption{Detection of Latitude and Longitude of Detected City.}
	\label{fig:fig10}
\end{figure}

\subsection{JSON Data for Visualization}
To give geolocation parameters to Google Map, We convert our data in JSON so that map can take latitude and longitude from there. Figure 12 shows the JSON file with all detected disease, their location name, latitude, longitude, and Data-base ID for easy tracking and source verification. 

\begin{figure}
	\centering
 \includegraphics[width=0.75\textwidth]{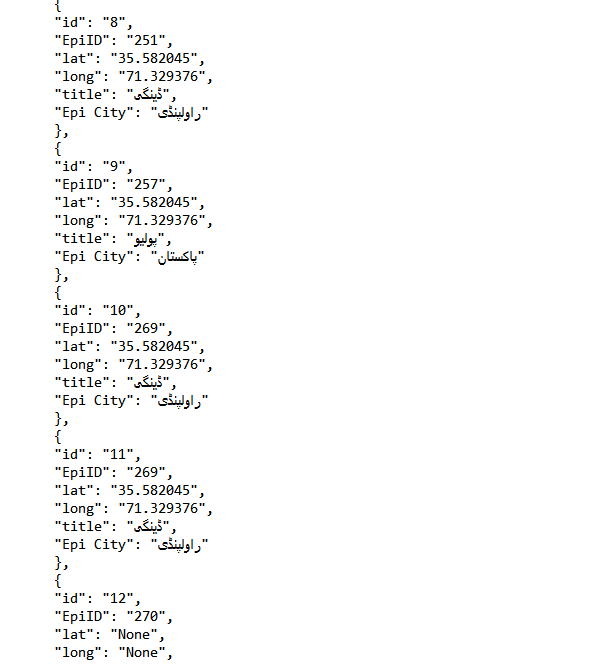}	
	\caption{JSON File.}
	\label{fig:fig11}
\end{figure}
\subsection{Google Map Visualization}
Finally, all the epidemic outbreaks and locations are shown in the Google Map with marker feature which is used to give more clarity on a map, sometime also refers as clustering, when many diseases is spread in nearby areas then clustering/ marker feature marked that area with a single marker and shows numeric to give more clarity shows in figure 13. 

\begin{figure}
	\centering
 \includegraphics[width=0.75\textwidth]{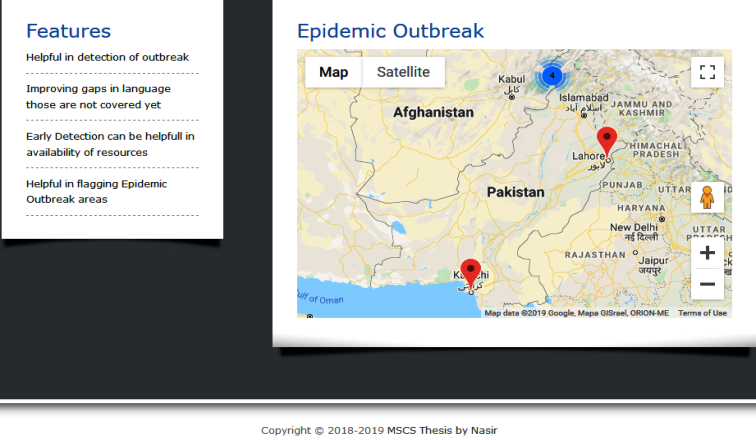}	
	\caption{Google Map with Clustering Marker.}
	\label{fig:fig12}
\end{figure}

\subsection{Removal of Diacritics:}
Not all the RSS feed has diacritics but very few, So identify it with Expert Identi-fication and figure it out that 55 RSS feeds which have diacritics. Now we analy-sis our system performance with Expert Identification, and In Table 1, we can see how many accuracies, precision, recall and F-Score will obtain

\begin{table}[h]
\centering
\caption{{Performance Evaluation of Diacritics Removal}}
\label{tab:tab1}
\begin{tabular}{cccccccccccccc}
\toprule
\multicolumn{2}{c}{Removal of Diacritics (Confusion Matrix)}                   \\
\cmidrule(r){1-2}
  \multicolumn{0}{c}{\textbf{Expert Identification}} &
  \multicolumn{2}{c}{\textbf{BIOPAK Flasher}} &
  \multicolumn{0}{c}{\textbf{Total}} &
\\

 \boldmath{}&\boldmath{Removed} & \boldmath{Not Removed} \\

\midrule
$-$Removed & $-$55 (TP) & $-$0(FN) & $-$55\\
Not Removed & $-$0(FP) & $-$0(TN) & $-$0 \\
$-$Total & $-$55 & 2 & {$-$55} \\

\bottomrule
\end{tabular}
\end{table}
\subsection{Removal of Punctuations: }
Every sentence ends up with (Full stop). Others are used in sentences. It is too much difficult and vague to quantify this performance test on the whole existing database, for this purpose we choose 50 RSS feeds that have (90) punctuations. Now we analysis our system performance with Expert Identification, and we can see in Table 2 that how many accuracy, precision, recall, and F-Score will obtain.

\begin{table}[h]
\centering
\caption{Performance Evaluation of Punctuation Removal}
\label{tab:tab2}
\begin{tabular}{cccccccccccccc}
\toprule
\multicolumn{2}{c}{Removal of Punctuation (Confusion Matrix)}                   \\
\cmidrule(r){1-2}
  \multicolumn{0}{c}{\textbf{Expert Identification}} &
  \multicolumn{2}{c}{\textbf{BIOPAK Flasher}} &
  \multicolumn{0}{c}{\textbf{Total}} &
\\

 \boldmath{}&\boldmath{Removed} & \boldmath{Not Removed} \\

\midrule
$-$Removed & $-$88 (TP) & $-$2(FN) & $-$90\\
Not Removed & $-$0(FP) & $-$0(TN) & $-$0 \\
$-$Total & $-$88 & 2 & {$-$90} \\
\bottomrule
\end{tabular}
\end{table}

\subsection{Tokenization Correctness:  }

A very important step in NLP \cite{mullen2018fast}, we have 301 RSS Feed for evaluation purposes. Now we analysis our system performance with Expert Identification, and see how much accuracies, precision, recall, and F-Score will obtain shows in Table 3. 

\begin{table}[h]
\centering
\caption{Tokenization Correctness  (Confusion Matrix)}
\label{tab:tab3}
\begin{tabular}{cccccccccccccc}
\toprule
\multicolumn{2}{c}{Removal of Diacritics (Confusion Matrix)}                   \\
\cmidrule(r){1-2}
  \multicolumn{0}{c}{\textbf{Expert Identification}} &
  \multicolumn{2}{c}{\textbf{BIOPAK Flasher}} &
  \multicolumn{0}{c}{\textbf{Total}} &
\\
 \boldmath{}&\boldmath{Tokenize} & \boldmath{Not Tokenize} \\

\midrule
$-$Tokenize & $-$273 (TP) & $-$16(FN) & $-$289\\
Not Tokenize & $-$11(FP) & $-$1(TN) & $-$12 \\
$-$Total & $-$284 & 17 & {$-$301} \\
\bottomrule
\end{tabular}
\end{table}
\subsection{Overall System Accuracy}

We conclude all the results and find the overall performance of our system. Starting from first layer shows in Table 4:

\begin{table}
	\caption{In terms of overall performance of BIOPAK Flasher, it performed above 80\% for almost all stages.}
	\centering
	\begin{tabular}{lllllll}
		\toprule
		\cmidrule(r){1-4}
		BIOPAK Flasher Stages    & Accuracy     & Precision  & Recall  & F- Score\\
		\midrule
		Removal of Diacritics & 100 \% & 100 \%   & 100 \%   & 100 \%    \\
		Removal of Punctuation    & 97.8\%	&  100\%	& 97.8\%	& 98.9 \%    \\
		Tokenization Correctness    & 91\%	  & 96.1\%	& 94.5\%	& 95.3\%  \\
Stop Word Removal   & 89.6\% &	92.6\%	 &89.8\% &	91.2\%  \\
Epidemic outbreak detection   & 93.9\%	 &97.4\% &	95.7\% &	96.5\% \\
City detection    & 96.4\%	 &97.8\%	&92.3\%	&98.1\%\\
City Lat Long Detection  & 99.4\% &	100\% &	99.4\%	 &99.7\%  \\
Overall BIOPAK Flasher & 95.4 \%	 &97.7\%	 &95.6\%	 &97.1\%  \\
		\bottomrule
	\end{tabular}
	\label{tab:table}
\end{table}
Pathogen \cite{lazcka2007pathogen} is the main cause of the epidemic outbreaks. These pathogens are required objects to survive and can spread from human to human. These pathogens can also spread through trade. It is necessary to stop these types of the pathogen to spread more and confined them. A carrier of such pathogen may cause sources of spread for others if travel other places. To escape and confined this outbreak, it is necessary to stop other people to visit that place or stop trade with that place, howev-er, if necessary, to continue relations then proper vaccination will be carried out. 

\subsection{Conclusion}

Our research points out the areas where the epidemic outbreak has occurred using Urdu language. To know such type of information, many actors are playing a vital role e.g., RSS Feed, Twitter, Social media, Printed Media, e-paper and News channels, etc. These resources are available in printed form (hard copy) or the e-electronic form (soft copy) in English and Urdu. To get news of the epidemic outbreaks, we choose the RSS feed in Urdu as a resource.\\
Urdu language is ignored and not used for epidemic detection. Many web sites are working, but most of the sites are in English. Whereas, Urdu is widely spoken language in Asia and also the national language of Pakistan. Majority people can better understand if the information is spread in Urdu. Very few media groups are maintained their RSS feeds in Urdu, it may be increased with the passage of time. \\
Our proposed model is using state of the art technologies of NLP for the ex-traction of data from these feeds. To implement this, we used the different  NLP li-braries. Some libraries have good support for Urdu but some have less support, which creates a problem for us. There is no full matured library is available for the NLP tasks on Urdu language. Every library has basic level support for Urdu, restriction of these libraries lies on the minimum level of Urdu data corpus for training different models. However, we extracted the epidemic diseases and cities from these RSS feed with the help of these libraries and some custom developed algorithm. \\
We divided our model into four layers, so that maintenance, troubleshoot-ing, and updating is so easy, we can add, modify or delete any layer.  The first layer is used to acquire RSS feed from RSS provider and saved it into the database. RSS feed duplication from same or different sources removed. In the second layer, we performed NLP techniques to filter the data, we take a single RSS feed’s title, removed diacritics, and punctuations in the very first stage. After these basic filtering, we to-kenized the data and removed stop words. At the end of the second layer, we have filtered unnecessary data, now it is easy to perform the next steps with great perfor-mance. Our third layer is used to extract epidemic diseases and their area of occur-rences. For this purpose, we developed three customized dictionaries with the en-hanced features of correction, and updating. These dictionaries have epidemic dis-ease ontology, Pakistani cities in Urdu, and their geo locations. These dictionaries are used as a reference, if any epidemic pathogen is detected after deployment of our system, then it is very easy to include this pathogen in our database. On detection of epidemic disease, our system then detect city from feed, if city is not found here then move our pointer to description portion for finding occurrence area. After finding the city, it is necessary to get its geo location which is required by Google Map to display occurrence areas. In the fourth layer, we visualized the disease on Google Map. The Map will be too messy if we display all the information there, to clutter down and give better visualization, we used the marker technique. In this layer, we also stored detected disease, city, and geo location in a database for displaying it up to 90 days. \v
We checked the performance of each layer with accuracy, precision, recall, and F-score. The overall system accuracy is 95.4\%, Precision 97.7\%, Recall 95.6\% and F-score is 97.1\%. In terms of the performance of BIO PAK Flasher, it performed above 80\% for almost all stages. We also give a comparison of different tools and libraries, which are available to perform said task. BIOPAK Flasher also gives re-source evidence, if any viewer is required to see the full news, our system can direct him to the concerned web page for a detail study. Our research gives a breakthrough in the field of Urdu language when working with online actors. However, more work and efforts are needed to mature Urdu NLP.

\bibliographystyle{unsrtnat}
\bibliography{references}  






\end{document}